\documentclass[prl,twocolumn,preprintnumbers,amsmath,amssymb]{revtex4}

\usepackage{graphicx}
\usepackage{ifthen}
\usepackage{ifpdf}
\usepackage{color}

\usepackage{latexsym}
\usepackage{amssymb}
\usepackage{bm}

\usepackage{tikz}
\definecolor{labelkey}{cmyk}{.4,.2,0,0}
\definecolor{Blue}{rgb}{0.00, 0.00, 1.00}
\definecolor{Red}{rgb}{1.00, 0.00, 0.00}
\definecolor{Green}{rgb}{0.40, 0.80, 0.60}
\definecolor{Op}{rgb}{1.00, 0.50, 0.00}

\newcommand{\blue}{\color{Blue}}

\def\be{\begin{equation}}
\def\ee{\end{equation}}

\def\bea{\begin{eqnarray}}
\def\eea{\end{eqnarray}}

\def\cb{\bar{c}}

\def\cb{\bar{c}}

\newcommand{\beq}[1]{\begin{eqnarray}\ifthenelse{#1=-1}{\nonumber}
{\ifthenelse{#1=0}{}{\label{e#1}}}}
\newcommand{\eeq}{\end{eqnarray}}
\newcommand{\hide}[1]{}

\begin{document}

\title{From the sine--Gordon field theory to the Kardar--Parisi--Zhang growth equation}
\author{Pasquale Calabrese{$^1$}, M\'arton Kormos{$^2$} and Pierre Le Doussal{$^3$} }

{\affiliation{{$^1$} Dipartimento di Fisica dell'Universit\`a di Pisa and INFN, 56127 Pisa, Italy}
{\affiliation{{$^2$} 
2 MTA-BME ÒMomentumÓ Statistical Field Theory Research Group, 1111 Budapest, Budafoki \'ut 8, Hungary}
 \affiliation{{$^3$} CNRS-Laboratoire de
Physique Th{\'e}orique de l'Ecole Normale Sup{\'e}rieure, 24 rue
Lhomond,75231 Cedex 05, Paris France.}

\begin{abstract}
We unveil a remarkable connection between the sine--Gordon quantum field theory and the Kardar--Parisi--Zhang (KPZ) growth equation. 
We find that the non-relativistic limit of the two point correlation function of the sine--Gordon theory is related to the generating function 
of the height distribution of the KPZ field with droplet initial conditions, i.e.\ the directed polymer free energy with two endpoints fixed. 
As shown recently, the latter can be expressed as a Fredholm determinant which in the large time separation limit converges to
the GUE Tracy--Widom cumulative distribution.  Possible applications and extensions are discussed. 
\end{abstract}


\maketitle

\noindent The sine--Gordon (sG) model and its cousin, the sinh--Gordon (shG) model, are paradigmatic integrable quantum field theories 
with countless applications in condensed matter physics (see e.g. \cite{ek-rev} as a review).
A lot is known in particular about their spectra in terms of many-particle scattering states and about matrix elements of local operators, 
the so-called form factors \cite{s-book,m-book}. This allowed  very striking predictions for experimentally relevant 
systems (such as spin chains and ladders \cite{ek-rev}) which in the scaling limit are described by these massive field theories. 
While the shG model contains only a single type of particle of mass $M$, the 
excitation spectrum of the sG model exhibits solitons as well as ``breathers'' $B_m$ that can be viewed as bound states 
of $m$ `particles' or, alternatively, as soliton-antisoliton bound states. 
The two models are also related by  an analytic continuation of the coupling constant.

As a recent experimentally relevant application, the non-relativistic limit (NRL) of the shG model was considered in Ref. \cite{Kormos1}, where it was shown  that in a double scaling limit (i.e.\ taking a NRL while taking the shG coupling to zero) 
the Lieb--Liniger (LL) model \cite{ll} is recovered with {\it repulsive} interactions.  
This procedure allowed the analytic calculation of some previously unknown local expectation values \cite{Kormos1,adilet}.
Taking the same double scaling limit of the sG model instead gives the Lieb--Liniger (LL) model \cite{ll} 
with {\it attractive} interactions, and indeed it is known that the scattering phases of the two models do coincide \cite{cc-07}.
However, this procedure has not yet been explored to obtain expectation values of measurable observables.

In an a priori completely different context, there has been much progress in finding
exact solutions to the 1D Kardar--Parisi--Zhang (KPZ) equation 
\cite{KPZ,we,dotsenko,spohnKPZEdge,corwinDP,we-flat,quastel-prep,SasamotoStationary,dg-12,ProlhacNumerics,flatnumerics,d-13,psn-11,ld-14,cl-14}.
Some of these approaches used the mapping onto the directed
polymer (DP) model, and from there, using the replica method, onto the LL model of bosons with attractive interactions \cite{kardareplica}.
These bosons form bound states called strings in the large system size limit  \cite{m-65},
and summation over these string states has allowed for the calculation of the probability distribution function (PDF) 
of the KPZ height field for various initial conditions 
\cite{we,dotsenko,we-flat,SasamotoStationary,dg-12}. 
Some of these predictions have been tested experimentally \cite{exp4}. 

The aim of this paper is to unveil a connection between the KPZ or DP models and the NRL of the sG quantum field theory
which arises because, as we mentioned above, in the NRL the sG model reproduces the {\it attractive} LL model. 
While $m$, characterizing the breathers, is a bounded integer for sG, 
it becomes unbounded in the NRL and reproduces the bound states of the LL model, the so-called string states.
Working out the details, we show that the connection is remarkably 
simple and, in particular, we find that the two point correlation function of exponential fields (vertex operators) in 
the sG model encodes the information of the PDF of the height field in KPZ.

{\it KPZ equation and Lieb--Liniger model:} Let us recall some facts about the KPZ equation in one space dimension. 
It describes the stochastic growth of an interface of height $h(x,t)$, as a function of time as
\bea 
\label{kpzeq2}
&& \partial_t h =  \nabla^2 h + (\nabla h)^2 + \eta
\eea
in dimensionless units, in the presence of white noise $\overline{\eta(x,t) \eta(x',t')}=2 \bar c \delta(x-x') \delta(t-t')$.
Here we focus on the so-called droplet, or sharp-wedge, initial condition, 
where $h(x,t=0)=-w |x|$ with $w \to +\infty$. The height field $h(x,t)$ can be written as $h=\ln Z$ where $Z$ is the
partition sum of a fixed endpoint DP, i.e.\ paths in the $x,t$ plane with endpoints fixed at $(0,0)$
and $(x,t)$, directed along the time direction. In the KPZ context one introduces the generating function
\bea
&& g(u) = \sum_{n=0}^\infty \frac{(-u)^n}{n!}  \overline{Z^n} \,,
\eea
which is a series representation for the Laplace transform $\overline{e^{-u Z}}$ of the PDF of Z. 
In the large time limit the (one-point) fluctuations of $h(x,t)$ are expected
to grow as $t^{1/3}$. 

To study this limit it is convenient to write 
$h(x=0,t) \simeq v t + 2^{2/3} \lambda \chi(t)$ where $\lambda=(\bar c^2 t/4)^{1/3}$ and $\chi(t)$ is an $O(1)$ random variable. 
We also define $u=e^{-\lambda s}$, since then $\tilde g(s) = g(u=e^{-\lambda s}) \to \text{Prob}(\chi(t) < s)$
for $\lambda \to \infty$, hence the generating function $\tilde g(s)$ is directly the cumulative distribution function (CDF)
of the height. 
The important finding of Refs. \cite{we,dotsenko} is that $\tilde g(s)$ can be expressed as
a Fredholm determinant at all times (see below) and that it converges for large time to $F_2(s)$, the GUE Tracy Widom CDF 
\cite{TW1994}. These results were obtained exploiting the property that the moments $\overline{Z^n} $
can be written
as the diagonal propagator of an $n$-particle Bose gas  with the Hamiltonian
\bea \label{LLmodel} 
H_\text{LL} = \frac1{2M}\sum_{\alpha=1}^n \partial_{x_\alpha}^2 - 2 \bar c \sum_{\alpha < \beta } \delta(x_\alpha - x_\beta)\,, 
\eea 
i.e.\ the attractive LL model which is integrable via the Bethe ansatz \cite{KBI}. 
More precisely, $\overline{Z^n} $ reads
\be
 \overline{Z^n} =\langle \vec x_0  |e^{-t H_\text{LL}}| \vec x_0 \rangle\, 
 = \sum_{\mu_n} \frac{|\langle \vec x_0 |\mu_n\rangle|^2}{||\mu_n ||^2} e^{-t E_{\mu_n}}\,,
\label{mom}
\ee
where $\mu_n$ are the $n$-boson eigenstates of $H_\text{LL}$ with eigenenergies $E_{\mu_n}$,
 $ \vec x_0=( x_0 \dots x_0)$, 
so $\langle \vec x_0 |\mu_n\rangle$ is the real space wave function evaluated 
at coinciding particle positions in $x_0$. Evaluating this sum over Bethe states 
in the infinite system size limit is not trivial and was performed in Refs. \cite{we,dotsenko}. Here we show
that the same sum can be retrieved rather simply, directly in the infinite system limit, from known results 
in the sG field theory, without the need to manipulate Bethe states. 
Besides its interest for the LL physics, it also raises hope that other
interesting quantities in KPZ could be calculated using the sG model.

{\it Sine--Gordon model}: 
The sG model is a relativistically invariant integrable quantum field theory in
($1+1$) dimension defined by the {\it Euclidean} (imaginary time) Lagrangian density
\bea
\mathcal{L}_{\rm sG} =  \frac{1}{2 c_l^2} (\partial_t \phi)^2 + \frac{1}{2} (\partial_x \phi)^2 - \frac{m_0^2 c_l^2}{\beta^2} (\cos(\beta \phi)-1)\,,
\label{sGlagr}
\eea 
where $\phi(x,t)$ is a real scalar field,  $c_l$ is the speed of light, $m_0$ the bare mass, 
$\beta $ the coupling constant,  and we set $\hbar=1$. 
The  renormalized (physical) coupling constant and mass are 
\bea
 \hat \alpha &= &\frac{c_l \beta^2}{8 \pi - c_l \beta^2} \,,\\
 M^2&=&m_0^2 \frac{\sin(\pi \hat \alpha)}{\pi \hat \alpha}\,.
\eea
The spectrum of the theory contains several kinds of particles. 
The `fundamental'  one is called 1-breather which has relativistic dispersion
relation with energy and momentum 
\be
E(\theta) = M c_l^2 \cosh(\theta)\,,\qquad p(\theta)= M c_l \sinh(\theta)\,,
\ee
where $\theta$ is the rapidity. 
The spectrum also contains solitons as well as $m$-breathers which are bound states of particles.
We will need below the dispersion relation of the $m$-breather. Its total energy and momentum
as a function of its center of mass rapidity $\theta$ can be obtained by introducing 
$\theta^a= \theta - \frac{2 a - 1 - m}{2} i \pi \hat \alpha$ and writing 
\bea
&& E_m(\theta)= M c_l^2 \sum_{a=1}^m \cosh(\theta^a) = M_m c_l^2 \cosh(\theta) \label{Em}\,, \\
&& P_m(\theta) = M c_l \sum_{a=1}^m \sinh(\theta^a) = M_m c_l \sinh(\theta)\, ,
\eea
leading to the same dispersion relation as for 1-breathers but with a 
different mass $M_m$ \cite{DHN}
\bea
M_m = M \frac{\sin( m \pi \hat \alpha/2)}{\sin(  \pi \hat \alpha/2)}\,.  
\label{Mm} 
\eea 
Note that the particle content $m$ of the breather is bounded in the relativistic sG model,
namely $m_\text{max} = \lfloor1/\hat \alpha\rfloor$. 
Due to its integrability, the sG model supports diffractionless factorized scattering and the exact 2-particle S-matrix is known \cite{ZZ,m-book}. 
Based on this, matrix elements of exponential vertex operators between scattering states, i.e.\ form factors have been computed exactly \cite{karowski, s-book, lukyanov}.

{\it Double non-relativistic limit and LL model:} Here we are interested in the (double) NRL defined as \cite{Kormos1}
\be
 c_l \to + \infty\,, \quad \beta \to 0\,, \quad \beta c_l = 4 \sqrt{\bar c} \,,
\ee 
with $\cb$ fixed and finite. Hence  the renormalized coupling tends to zero as
\bea 
\hat \alpha  \simeq \frac{c_l \beta^2}{8 \pi} = \frac{2 \bar c}{\pi c_l} \to 0\,.
\eea
In this limit the  dispersion relation of the particles become{\blue s} non-relativistic as
\be
 p  \simeq M c_l \theta \equiv  \lambda\,, \qquad
 E  \simeq  M c_l^2 + \frac{\lambda^2}{2 M}\,,
\ee 
where $\lambda$ is the usual (non-relativistic) rapidity, i.e.\ the quasi-momentum.
Importantly, this double limit establishes a connection between the exact S-matrices and the form factors of the two models.
At the level of the fields the correspondence can be written in the double limit as
\bea
\phi(x,t) = \frac{1}{\sqrt{2 m_0}} \left[  \Psi(x,t) e^{- m_0 c_l^2 t } + \Psi^\dagger(x,t) e^{m_0 c_l^2 t } \right] \,,
\label{phivspsi}
\eea 
where $\phi$ is the sG field and 
$\Psi^\dagger$ creates a non-relativistic particle (i.e.\ the LL boson). 
Plugging (\ref{phivspsi}) into (\ref{sGlagr}) leads, upon expansion and neglecting highly oscillating terms, 
to the non-linear Schr\"odinger Hamiltonian 
\begin{equation}
H_{\rm LL} = \int dx \left( \frac1{2M}\nabla \Psi^\dag\nabla\Psi - \cb \Psi^\dag\Psi^\dag\Psi\Psi\right) \,,
 \label{rhodef}
\end{equation}
which is the second quantized form of the attractive LL Hamiltonian (\ref{LLmodel}).
This limit procedure was first shown for the shG model in Ref. \cite{Kormos1}, leading to the repulsive LL model with 
interaction parameter $c=- \bar c>0$. 
Here the same method shows that $\bar c>0$ emerges as the coupling constant of the attractive LL model. 
In fact, at the level of the single particle states a lot can be deduced by analytical continuation 
from shG to sG ($\beta \to i \beta$). This technique was used to study a highly excited gas-like state of the attractive LL model, the super Tonks-Girardeau gas \cite{sTG}. However,  the many-particle states in general are quite different in the repulsive ($\bar c<0$) and attractive ($\bar c>0$) cases. Consider the energy of the sG $m$-breather (\ref{Em}-\ref{Mm}) in the double NRL:
\bea
E_m(\theta) \simeq M m c_l^2 +  \frac{\bar c^2}{24M}  (m - m^3)  + m \frac{p^2}{2 M} \,,
\eea 
while the momentum is $P_m(\theta) = m p$, where we have scaled $\theta=p/(M c_l)$ and neglected terms $O(1/c_l)$. 
Apart from the rest energy, these are exactly the total energies and momenta of the  $m$-string states of the LL model
in the limit of infinite system size \cite{m-65}. 
The correspondence goes further and indeed also the scattering phases coincide, as pointed out in \cite{cc-07}.
Finally, in the NRL the mass of the solitons/antisolitons $M_s=8 m_0^2/\beta^2$ \cite{m-book} diverges, 
hence they disappear from the spectrum and we can neglect them. 
We are then left with an infinite number of breather modes as $\hat \alpha \to 0$ corresponding to the LL strings. 
The form factors of the LL model can also be obtained through the NRL from the breather form factors of the sG model, 
paralleling the calculation for the shG model \cite{Kormos1, kmt-09}.
In the remainder of the paper we will set $M=1/2$ by a choice of units,
as customary in the LL model. 

{\it  From sG correlations to the KPZ/DP model:} Let us consider  the two-point correlator
of the exponential field in the Euclidean sG model (i.e.\ in imaginary time) 
\be 
 G(\tilde k,t) = \langle0| e^{ i \tilde k \phi(0,t) } e^{-  i \tilde k \phi(0,0) } |0\rangle\,,
 \label{ratio}
\ee
as well as the reduced correlation
\footnote{ In our normalization indeed $|\langle e^{i \tilde k \phi}\rangle|^2=1$ and taking the ratio (\ref{ratio}) is superfluous. 
However, this ratio is independent of normalizations and conventions and the results for this quantity are fully general.}
\bea 
&& \tilde G(\tilde k,t) = G(\tilde k,t)/{|\langle e^{i \tilde k \phi}\rangle|^2} \,,
\eea
which is defined to equal unity at $t \to \infty$. Here and below we denote
the vacuum expectation value of the exponential field, 
as $\langle e^{i \tilde k \phi}\rangle=\langle0| e^{ i \tilde k \phi(0,0) }|0\rangle$.

The Lehmann formula \cite{ek-rev,m-book,s-book} expresses 
quite generally such a ground state expectation value (here in the vacuum) in terms of the
form factors of the excitations of the theory. At this stage we do not yet consider 
the NRL but, for simplicity, we ignore the solitons states (which will be justified only in that limit).
It thus takes the form of a sum over states with arbitrary number $n_s$ of breathers
\begin{widetext}
\be
G(\tilde k,t) \simeq \sum_{n_s=0}^\infty \frac{1}{n_s!} \prod_{j=1}^{n_s} \sum_{m_j=1}^{m_{max}} 
\int \frac{d\theta_1}{2 \pi}\dots  \frac{d\theta_{n_s}}{2 \pi} 
|\langle 0 | e^{ i\tilde k \phi(0,0) } |B_{m_1}(\theta_1) \dots B_{m_{n_s}}(\theta_{n_s}) \rangle|^2 
e^{-  \sum_{j=1}^{n_s} E_{m_j}(\theta_j)  |t| }\,, 
\label{G}
\ee
\end{widetext}
where each breather  of type $m_j$ has rapidity $\theta_j$ and particle content $m_j$.

The form factors of the breathers can be obtained from those of the particles $B_1(\theta)$. Hence
let us start with the particle states, i.e.\ we temporarily restrict to $m_j=1$ in (\ref{G}). 
Their form factors are known in sG \cite{takacs} and can be obtained from the shG ones \cite{koubek} 
via analytical continuation to imaginary coupling as
\begin{multline}
 F_n^{k}(\theta) = \langle 0|e^{i k \beta \phi} |\theta_1\dots \theta_n \rangle =  \langle e^{i k \beta \phi}\rangle \frac{\sin(k \pi \hat \alpha)}{\sin(\pi \hat \alpha)}  (2 i)^n  \\
 \times \Big(\frac{\sin(\pi \hat \alpha)}{F_{min}(i \pi)}\Big)^{n/2} \det M_n(k) \prod_{j < l} \frac{F_{min}(\theta_j-\theta_l)}{e^{\theta_j}+ e^{\theta_l}}\,, 
 \label{FF}
\end{multline}
where $|\theta_1\dots\theta_n \rangle=|B_{1}(\theta_1) \dots B_{1}(\theta_{n}) \rangle$. 
Here we introduced the matrix (with $j,l =1,\dots n-1$)
\be
[M_n(k)]_{j,l} = \frac{\sin[(j-l+k) \pi \hat \alpha]}{\sin(\pi \hat \alpha)}   \sigma^{(n)}_{2 j - l}\,, 
\ee
for $n \geq 2$,
where $\sigma^{(n)}_{j}$ is the $j$-th elementary symmetric polynomial of the variables $\{e^{\theta_1},\dots,e^{\theta_n}\},$ 
and ${\rm det} M_1(k)=1$. The minimal form factor is given by
\be
F_\text{min}(\theta)={\cal N} e^{-4\int_0^\infty\frac{d t}t
\frac{\sinh(\frac{t}2\,\hat\alpha)\sinh(\frac{t}2(1+\hat\alpha))}
     {\sinh(t)\cosh(\frac{t}2)}
     \sin^2\left(\frac{t(i \pi-\theta)}{2\pi}\right)}\,,
\label{FMINSHG}
\ee
where ${\cal
N}=F_\text{min}(i\pi)=    e^{\frac1\pi\int_0^{\pi\hat\alpha}d t\,\frac{t}{\sin(t)}} / \cos\left(\frac{\pi\hat\alpha}2\right)$. 
The minimal form factor satisfies the exact relation \cite{koubek}
\bea \label{productrel} 
F_{min}( i \pi + \theta) F_{min}(  \theta) =  \frac{\sinh \theta}{- \sinh(i \hat \alpha \pi) + \sinh \theta} \,.
\eea

We now carefully take the (double) non-relativistic limit of the form factors. 
To obtain a non-trivial limit we need  to let $k\to\infty$ while $\beta \to0$ with
$\tilde k= k\beta$ fixed and finite. 
We also scale the rapidities as $\theta_j = 2 p_j/c_l$ (we recall $M=1/2$). 
Given that ${\cal N} \to 1$ as $\hat \alpha \to 0$, using Eq. (\ref{productrel}) we have in the NRL
\bea
F_{min}(\theta)  
\to \frac{\theta}{- i \hat \alpha \pi + \theta} \,.
\eea
The most complicated term in the form factor (\ref{FF}) is $\det M_n$ which immensely simplifies in the NRL 
because  one can neglect the $i-j\ll k$ in the matrix $M_n$ to get
\be
  \det \Big[M_n(\tilde k/\beta)\Big]   \simeq  \Big(\frac{\sin(k \pi \hat \alpha)}{\sin(\pi \hat \alpha)}\Big)^{n-1} {\rm det}  \sigma^{(n)}_{2 j - l} \,,
\ee
and ${\rm det}  \sigma^{(n)}_{2 j - l} \to 2^{n(n-1)/2}$ in the NRL. 
Finally, in the NRL we also have the simple relation 
%
\be
 \frac{\sin(k \pi \hat \alpha)}{\sin(\pi \hat \alpha)} \to \frac{\sin(\frac{\sqrt{\bar c}}{2} \tilde k)}{\pi \hat \alpha}  \,.
\ee
All these relations lead for the most complicated parts of the form factor to the remarkably simple limit
\begin{multline}
\frac{\sin(k \pi \hat \alpha)}{\sin(\pi \hat \alpha)}  2^n  
  \Big(\frac{\sin(\pi \hat \alpha)}{F_{min}(i \pi)}\Big)^{n/2} \det M_n(k) \prod_{j < l} \frac{1}{e^{\theta_j} + e^{\theta_l}}\\
  \to  \Big(\frac{c_l}2\Big)^{n/2}  \Big[ \frac{2}{\sqrt{\bar c} } \sin( \frac{\sqrt{\bar c}}{2}  \tilde k ) \Big]^{n}\,. 
\end{multline}
Plugging this in (\ref{G}) and taking into account the Jacobian of the variable change from $\theta$'s to $p$'s, we find for the $m_j=1$ contribution
\begin{multline}
 G(\tilde k,t) \simeq |\langle e^{i \tilde k \phi}\rangle|^2 \sum_{n=0}^\infty \frac{1}{n!}  
\Big[ \frac{2}{\sqrt{\bar c} } \sin( \frac{\sqrt{\bar c}}{2}  \tilde k ) \Big]^{2 n}  
 e^{- n M c_l^2 |t| }  \\
  \int \frac{dp_1}{2 \pi} \dots  \frac{dp_n}{2 \pi}  \prod_{j < l} 
\frac{(p_j-p_l)^2}{\cb^2 + (p_j-p_l)^2} e^{- \sum_j p_j^2 |t|} \,,
\end{multline}
which is a sum of positive contributions, as it should, since it comes from the Lehmann formula.

Let us now generalize the derivation to arbitrary $m$-breather states.
Fusion relations relate the form factors of breathers to the ones of $n$ particles $F_n$ as follows. 
Let us recall that the rapidities can be written as $\theta_j^a= \theta_j - \frac{2 a - 1 - m_j}{2} i \pi \hat \alpha$. 
Then, recalling $n=\sum_j m_j$, we have \cite{takacs} 
\begin{multline}
 \langle 0| e^{{\blue i} k \beta \phi} | B_{m_1}(\theta_1)\dots B_{m_{n_s}}(\theta_{n_s}) \rangle =\\
\prod_{j=1}^{n_s} \gamma_{m_j} F_{n}(\{ \theta_1^{a_1} \}_{a_1=1,\dots m_1},\dots\{\theta_{n_s}^{a_{n_s}}\}_{a_{n_s}=1,\dots m_{n_s}})\,,
\end{multline}
where the NRL of $\gamma_m$ given in \cite{takacs} is 
\bea
\gamma_m = (\pi \hat \alpha)^{(m-1)/2} \Gamma[m] \sqrt{m} \,.
\eea 
Note that $\theta_j^a \to \frac{\pi \hat \alpha}{\bar c} [p_j - \frac{i \bar c}{2} (2 a - 1- m_j) ]$
which coincide with the string rapidities \cite{m-65}. 

In taking the NRL of the form factor $F_n$, only the term with $F_{min}$ changes compared to the previous case, leading to
\be
 \prod_{1 \leq j < l < n} | F_{min}(\theta_j-\theta_l) |^2 = \Phi[p,m]  \prod_{j=1}^{n_s} | F[m_j] |^2 \,,
\ee
where
\begin{align}
 \Phi[p,m] &= \prod_{1 \leq j < l \leq n_s}
\frac{4 (p_i - p_j)^2 + \bar c^2 (m_i-m_j)^2}{4 (p_i - p_j)^2 + \bar c^2 (m_i+m_j)^2} \,,\\
 F[m] &= \prod_{1 \leq a < b \leq m} \frac{ b-a}{b-a+1} = \frac{1}{m!} \,.
\end{align}
Putting everything together we finally obtain
\begin{multline}
 G(\tilde k,t) \simeq |\langle e^{i \tilde k \phi}\rangle|^2 
 \sum_{n_s=0}^\infty  \frac{\bar c^{n- n_s}}{n_s!} 
 \prod_{j=1}^{n_s} \sum_{m_j=1}^{+\infty}   \Big[ \frac{2}{\sqrt{\bar c} } \sin( \frac{\sqrt{\bar c}}{2}  \tilde k ) \Big]^{2 m_j} \\
\prod_{j=1}^{n_s} \int \frac{dp_j}{2\pi m_j}  
e^{- m_j M c_l^2 t - \frac{\bar c^2}{12} (m_j^3-m_j) t - m_j p_j^2 t } \Phi[p,m]  \,.  
\label{Gktfin}
\end{multline}

Now we can compare this with the expression for the moments of the
partition sum in the KPZ/DP problem.
The calculation of the averaged moments (\ref{mom}) was performed in Ref. \cite{we} and found to take exactly the same
expression as above  (compare with Eq. (9) of Ref. \cite{we}).  Hence we find that the reduced correlation can be written as
\be
 \tilde G(\tilde k,t)  = \sum_{n=0}^\infty \frac{1}{n!}  
 \Big[ \frac{2}{\sqrt{\bar c} } \sin( \frac{\sqrt{\bar c}}{2}  \tilde k ) \Big]^{2 n}  \overline{Z^n} e^{- n M c_l^2 t } \,, 
 \label{rel1} 
\ee
showing that there is a 
relation between the two-point correlation in sG and 
the moments in the KPZ/DP problem.  Furthermore, the sG correlation
takes the same form as the KPZ/DP generating function, hence
one can also write
\be 
\tilde G(\tilde k,t)  = g(u)\,,
\ee 
where $u= - [ \frac{2}{\sqrt{\bar c} } \sin( \frac{\sqrt{\bar c}}{2}  \tilde k ) ]^2 e^{- M c_l^2 t }$.
Interestingly, by analytic continuation $i \tilde k \to \tilde k$ one also finds
\begin{multline}
 \frac{\langle 0| e^{ \tilde k (\phi(0,t) - \phi(0,0) ) } |0 \rangle}{\langle e^{\tilde k \phi}\rangle^2} =\\
 \sum_{n=0}^\infty \frac{(-1)^n}{n!}  \Big[ \frac{2}{\sqrt{\bar c} } \sinh( \frac{\sqrt{\bar c}}{2}  \tilde k ) \Big]^{2 n}  \overline{Z^n} e^{- n M c_l^2 t } 
 = g(u) \,, 
 \label{rel2} 
\end{multline}
where now $u=[ \frac{2}{\sqrt{\bar c} } \sinh( \frac{\sqrt{\bar c}}{2}  \tilde k ) ]^2 e^{-  M c_l^2 t } $
is a positive number,  thereby making the connection closer. Since the KPZ generating function obeys $0<g(u)<1$, it implies that 
both sides of the above equation are now positive numbers in the interval $[0,1]$, increasing with $\tilde k$. 
Note that although the operator in the left hand side of Eq. (\ref{rel2}) may not be formally defined in the sG field theory, 
in the Euclidean version it takes the meaning of a canonical statistical mechanics average, 
with a discretization and regularization at small and large scale (as one would do in a numerical simulation). 
While the two correlations in Eq. (\ref{rel2}) may be singular as the regularizations are removed, their ratio should be a well defined 
number in the interval $[0,1]$.

Having shown that the NRL of the sG correlation contains information about the
PDF of the KPZ field, we can ask whether a tighter physical connection exists. 
Let us recall the expression of the generating function as a Fredholm determinant (FD) obtained in
\cite{we}:
\be
\tilde g(s) = {\rm Det}[ 1 + P_0 K_s P_0 ] \,,
\ee
where $P_0$ is the projector on $[0,+\infty[$ and the kernel can be written as 
\be
K_s(v,v') = - \int \frac{d k}{2 \pi} dy Ai(y + k^2 + s + v+v') \frac{e^{\lambda y - i k (v-v')}}{1+ e^{\lambda y}} \,,
\ee
where $\lambda \sim t^{1/3}$ was defined above. Let us also recall that at large time this generating function $\tilde g(s)$ converges to $F_2(s)$, the GUE Tracy Widom CDF \cite{TW1994}. So it is interesting that this FD
contains the information about the precise time dependence of the
coefficients of each power of $e^{- M c_l^2 t }$ in the decay of the sG correlation function, 
these coefficients being proportional to the KPZ/DP moments $\overline{Z^n}$.
Note however, that because of these fast decaying exponentials, there is no point-wise convergence 
as $c_l \to \infty$ in (\ref{rel1}) and (\ref{rel2}) and, at this stage, there is no {\it direct} correspondence between a sG observable and 
KPZ generating function. An outstanding question is thus how far can this correspondence be pushed
and whether one can construct sG observables with an even closer relation to KPZ.

One last interesting point relates to the so-called moment problem in the continuum KPZ/DP problem, i.e.\
to the fact that the growth of the integer moments $\overline{Z^m}$ as a function of $m$ 
at fixed $t$ is too rapid (i.e.\ $\ln \overline{Z^m} \sim m^3 t$) to guarantee a unique solution for the PDF for $h=\ln Z$. 
While this is still an open question in the mathematics community, it is usually
circumvented by resorting to discrete models (such as TASEP, see e.g. \cite{spohnKPZEdge}) which do reproduce continuum KPZ in 
some limit and do not suffer from the same problem. 
We point out here that the relativistic sG theory can provide yet another interesting regularization of the moment problem 
because the number of breather types $ \lfloor1/\hat \alpha\rfloor$ is bounded until the NRL is taken.

{\it Overlaps and propagators in the LL model:} 
One can ask more precisely how can the sG theory retrieve more detailed information about the attractive LL model hereby
helping solve KPZ problems.
For example a related fundamental quantity (both for the LL model and the KPZ growth) 
is the overlap
\be
G(|\psi_1\rangle, |\psi_0\rangle ;t)=\langle \psi_1 |e^{-t H_\text{LL}}| \psi_0 \rangle
\ee
between more general initial $|\psi_0\rangle$ and final states $|\psi_1\rangle$. 
Examples are (i) KPZ with flat initial condition, which requires the overlap with an initial uniform state and whose solution was 
obtained in \cite{we-flat} (ii) the imaginary-time propagator for arbitrary positions \cite{ps-11,tw-08}. 
Here we have shown that in the NRL the diagonal propagator  can be retrieved as
\footnote{The case $G(|\vec x\rangle,|\vec y\rangle;t)$ with 
$\vec x=(x_0\dots x_0)$ and $\vec y=(y_0\dots y_0)$ is a trivial
generalization of the present formula, introducing the total momentum in 
Eq. (\ref{prop}), see e.g.  \cite{we-flat}.}
\be
\tilde G(\tilde k,t)|_{e^{- n M c_l^2 t }}  = \frac{1}{n!}  
 \Big[ \frac{2}{\sqrt{\bar c} } \sin( \frac{\sqrt{\bar c}}{2}  \tilde k ) \Big]^{2 n} G(|\vec x_0\rangle,|\vec x_0\rangle;t)\,. 
\label{prop} 
\ee 
It would be interesting to obtain more general overlaps via this sG correspondence which we leave for
future investigations. 

Relation \eqref{prop} can be understood in the following way. 
The diagonal propagator, $\langle\Psi^n(t){\Psi^\dagger}^n(0)\rangle$ can be extracted from the NRL of the sG correlator 
$\langle\phi^n(t)\phi^n(0)\rangle$ where the operator $\phi^n$ can be obtained naively from the $n$th order of the series expansion 
of $e^{ik\phi}$ in powers of $k$. 
However, it turns out that the $\phi^n$ operator obtained in this way has non-zero matrix elements between states having any number 
of particles. These form factors survive in the NRL meaning that in the limit one cannot recover the $\Psi^n$ operator. 
It turns out that certain linear combinations of different powers of $\phi$ lead to $\Psi^n$ in the NRL. 
Quite interestingly, these combinations in terms of $k$ give just the sine factor in Eq. \eqref{prop}, 
so this formula, quite miraculously, automatically takes care of this operator mixing.

{\it Conclusion}.
In this paper we have shown  yet another method to calculate the PDF for the KPZ growth equation with the narrow wedge 
initial condition exploiting the NRL limit of sG field theory. 
The obtained result fully agrees with previous derivations \cite{we,dotsenko,spohnKPZEdge,corwinDP}.
On the one hand this result is a useful check of our new method and of present and previous results, 
on the other hand this new correspondence raises the hope that it can be used to obtain yet unknown observables
for the KPZ growth equation, as well as provide a new interesting regularization of the problem.

{\it Acknowledgements}:
MK thanks G\'abor Tak\'acs for useful discussions. PC acknowledges the ERC  for financial  support under Starting Grant 279391 EDEQS. 
MK acknowledges financial support from the Marie Curie IIF Grant PIIF-GA-2012- 330076.
PLD acknowledges the hospitality of the 
Dipartimento di Fisica dell'Universit\`a di Pisa where much 
of this work was completed.

\end{document}